\documentclass[linenumbers]{aastex}

\usepackage{graphicx}
\usepackage{wrapfig}
\usepackage{natbib}

\newcommand{\be}{\begin{equation}}
\newcommand{\ee}{\end{equation}}
\newcommand{\nn}{\mbox{} \nonumber \\ \mbox{} }
\newcommand{\ba}{\begin{eqnarray}}
\newcommand{\ea}{\end{eqnarray}}
\newcommand{\om}{\omega}
\newcommand{\Alfven}{Alfv\'{e}n }

\newcommand{\E}{{\bf E}}

\newcommand\eg{{\it{e.g.\ }}}

\newcommand{\Bf}{{magnetic field}}

\newcommand{\Ef}{{electric  field}}

\newcommand{\NS}{neutron star}

\newcommand{\EM}{electromagnetic}

\newcommand{\ms}{magnetosphere}
\newcommand{\mss}{magnetospheres}

\newcommand{\LC}{light cylinder}
\newcommand{\Lf}{Lorentz factor}

\begin{document}

\title{Faraday conversion  in pair-symmetric    winds of  magnetars and  Fast Radio Bursts}
\author{Maxim Lyutikov}
\affil{
 Department of Physics and Astronomy, Purdue University, 
 525 Northwestern Avenue,
West Lafayette, IN, USA}

\begin{abstract}
We consider  propagation of polarization in the inner parts of  pair-symmetric   magnetar winds, close to the \LC. Pair plasmas in \Bf\ is birefringent, a $\propto  B^2$ effect. As    a result,  such plasmas work as  phase retarders: Stokes parameters follow a circular trajectory on the Poincare sphere. In the  highly magnetized regime, $\om, \, \om_p  \ll \om_B$,  the corresponding rotation rates are independent of the \Bf. A plasma screen with  dispersion measure DM $\sim 10^{-6}$  pc cm$^{-3}$ can induce large polarization changes, including large effective Rotation Measure (RM).  
The frequency scaling of the (generalized)  RM, $ \propto \lambda ^\alpha  $,  mimics  the conventional  RM with $\alpha =2$ for small phase shifts, but can be as small as  $\alpha =1$. In interpreting observations the  frequency scaling of polarization parameters should be fitted independently.
The model  offers explanations for  (i)  large circular  polarization component observed in FRBs,  with   right-left switching; (ii) large RM, with possible sign changes; (iii) time-depend variable polarization. Relatively  dense and  slow wind is needed - the corresponding effect in  regular  pulsars is small. 
\end{abstract}

\maketitle


\section{Polarization of FRBs: the challenges}
\label{intro}
Polarization properties of FRBs defy simple classification \cite{2019MNRAS.487.1191C,2019A&ARv..27....4P}, \citep[``some FRBs appear to be completely unpolarized, some show only circular polarization, some show only linear polarization, and some show both,''][]{2019A&ARv..27....4P}.  Understanding polarization behavior is the key to understanding FRBs.

Even in the sub-set of linearly polarized FRBs, there is no clear trend:
\begin{itemize} 
\item FRB 150807 \citep{2016Sci...354.1249R}  was nearly 80\% linearly polarized, but very small RM=12 at DM=266 (in usual astronomical unites); the average inferred magnetic field  $< B >= 5 \times  10^{-8}$  G.
\item FRB 110523 \citep{2015Natur.528..523M},  RM=186,  DM=623, $<B>=5\times10^{-7}$ G
\item FRB180301 \citep{2019MNRAS.486.3636P},  RM =  $3\times 10^3$, DM=522, $<B>=6\times10^{-6}$ G.
\item FRB121102 \citep{2018Natur.553..182M}  was 100\% linearly polarized, (varying!) RM = $10^5$, DM=559, $< B >= 2 \times 10^{-4}$  G. At the observed frequency of $\sim$  4.5 GHz this corresponds to the PA rotation by 360 radians; this is a model-independent quantity to  be explained (in a sense that a value of RM assumes a particular frequency scaling of the rotation of polarization).
\item The case of  FRB20190520B \citep{2022arXiv220308151D} is particularly interesting: it shows large fluctuations of linear polarization (from $\sim 15\%$ to $\sim 60\%$, 
 large fluctuations of circular polarization with changing sign ($\sim \pm 10\%$) and large fluctuations of RM, also with changing sign ($\pm 10^4$)
 \item no correspondingly large changes of DM are seen.
\end{itemize}

We make the following conclusion: polarization model should explain  not the average properties, but the extremes of the behavior. Models   should account for large variations in polarization properties, both between different sources, temporal variations in a given source, and unusual polarization behavior like in FRB180301 and FRB20190520B.

Fast temporal variation of RM seen  in FRB 121102 \citep{2018Natur.553..182M}  and  FRB20190520B \citep{2022arXiv220308151D}  are especially demanding, as this implies that the RM comes from a relatively compact region. This would normally  require  small and   {\it extremely  dense}  region,   yet no DM variations are seen. (In our model the region is small, but not dense.)

In this paper  we consider polarization propagation   effects in the near wind zone of the central magnetar, somewhat outside the \LC.  We demonstrate that the model naturally explains a broad range of polarization behavior. Switching of the signs of circular polarization, and of the (generalized) Rotation measure are especially noteworthy. 
We also mention related   papers by \cite{2019MNRAS.485L..78V} and  \cite{2019ApJ...876...74G} (in passing we note a minor error: their parameters $h$ and $g$ should be interchanged).

 Previously a number of works considered PA rotation  effect inside the pulsar \ms\  \cite{1979ApJ...229..348C,1986ApJ...303..280B,2000A&A...355.1168P,2010MNRAS.403..569W,2012MNRAS.425..814B}. Inside the pulsar \ms,  the PA  rotation and the generation of V component    are   suppressed by a  combination  of effects: (i)  for parallel propagation in symmetric pair plasma the  Faraday effect is absent; the contribution to the  PA rotation  comes either  from a  slight charge-disbalance, different {\Lf}s,  or from oblique propagation \citep{1991MNRAS.253..377K,1999JPlPh..62...65L}  - all these effects  producing weak contribution (due to small ``active'' density and/or small angle of propagation).  Relativistic motion of plasma  also reduces the effective plasma frame density, and stretching  of the corresponding time scale in the lab frame;  this is  effect is also  important in the present model.
In contrast to the \mss, in the near wind zone it  is  the {\it  total plasma density} that  contributes to the PA rotation, via the effects of birefringence.

\section{Polarization propagation in  birefringent symmetric pair plasma}

\subsection{Faraday and  Cotton-Mouton/Voigt effects}

Two somewhat different effects contribute to the changes  of polarization as the light propagates  in plasma: the Faraday effect and  the Cotton-Mouton/Voigt effects \citep{LLVIII}. Qualitatively, the Faraday effect is that a linearly polarized wave propagating along the \Bf\ can be decomposed into two circularly polarized waves. In the electron-ion plasma the  two circularly polarized waves have different phase velocity - their final addition leads to the rotation of the position angle (PA).
The rate of rotation of the polarization angle due to the Faraday effect is  \citep[][Eq. (4.6)]{1965ARA&A...3..297G}
 is
\be
 \frac{d\chi_F}{dl}=\frac{1}{2} \frac{\om}{c} (\Delta n)_c
\label{01} \ee
where $ (\Delta n)_c $ is the difference in the refractive index of two circularly polarized normal modes.
 This effect is linear in \Bf. It disappears in symmetric  pair plasma. 

The Cotton-Mouton/Voigt effects appear because
for oblique propagation (with respect to the \Bf), the two plasma modes, usually called  O (ordinary)  and X  (extraordinary), have different phase velocities. If the initial  wave had contribution from both X and O modes, the final addition of the retarded waves leads to elliptical polarization, hence  both to the rotation of the position angle of linearly polarized component, and to the appearance of circular polarization.  This effect is quadratic  in \Bf-- it appears both in symmetric and non-symmetric plasmas.

Following the tradition we call polarization transformation as Faraday Conversion (FC), with a clear understanding of the different origin of the circular component, as discussed above.

\subsection{Waves in symmetric  pair plasma}

\begin{figure}[h!]
\includegraphics[width=0.85\linewidth]{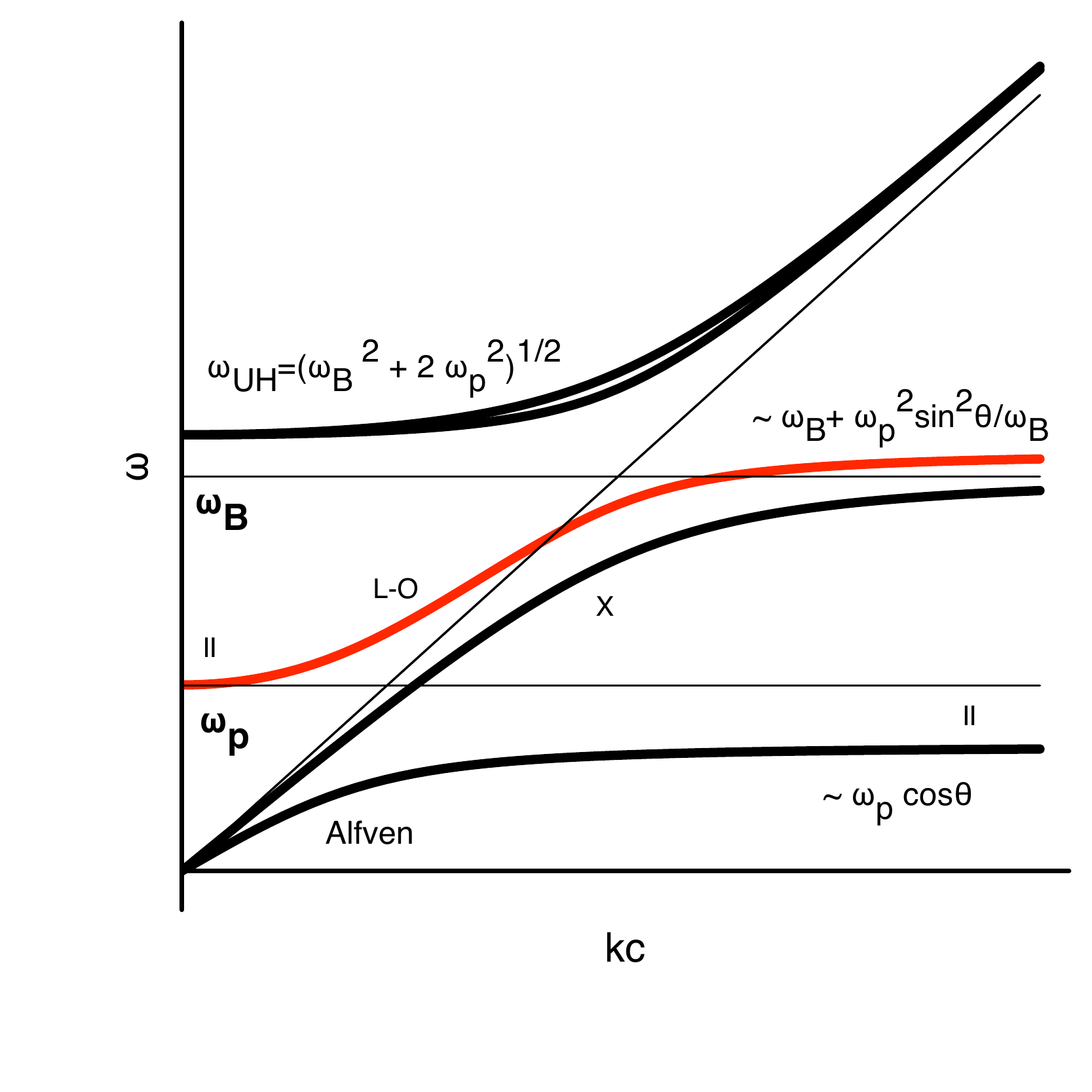}
\caption{Wave dispersions  $ \om(k) $ in pair plasma in strong \Bf, $\om_B \gg \om_p$, for oblique propagation. At low frequencies $\om \ll \om_B$ there are three modes labeled X (polarized orthogonally to   {\bf k} -{\bf B} plane), \Alfven and  O  (both polarized in the   {\bf k} -{\bf B} plane). The  O mode has a resonance at $\sim \om_B + \om_p^2 \sin^2 \theta/\om_B$ and cut-off at $\sqrt{2} \om_p$. The \Alfven mode has a resonance at  $\sim \sqrt{2} \om_p \cos \theta$. The sign $\parallel$ indicates locations where corresponding  waves are nearly longitudinally polarized. The two high frequency, $\om > \om_B$, waves with nearly identical dispersion have a cut-off at the upper hybrid frequency $\om_{UH}=\sqrt{\om_B^2 +2 \om_p^2}$,  \protect\citep{2007MNRAS.381.1190L} }
\label{1}
\end {figure}

Waves in pair plasma has been considered in a number of publications  \citep{AronsBarnard86,Kaz91,1999JPlPh..62...65L}  we follow \cite{2007MNRAS.381.1190L}. Let us consider the simplest case of cold plasma, in plasma frame. 
For  $e^\pm$ plasma in \Bf\ the  dispersion relation factorizes giving two modes: the
X mode  with the
electric vector perpendicular to the {\bf k-B} plane 
and two branches of the
longitudinal-transverse mode, which we will call 
 O
and Alfv\'{e}n waves,  with
 the electric vector in the {\bf k-B} plane \citep[][see Fig. \ref{1}]{AronsBarnard86}.
 X waves is a
subluminal  (for $\om < \om_B$)  transverse electromagnetic wave with a dispersion relation
 \be
n_X^2 = 1  - \frac{ 2 \om_p^2}{  \om^2 -\om_B^2}
\label{3} \ee
here $n= kc/\om$ is refractive index, $\om_B =e B/mc $ is cyclotron frequency,  $\om_p =\sqrt{ 4\pi n_{\pm} e^2/m}$ is a plasma frequency of each species (so that for pair plasma the total plasma frequency is $\sqrt{2} \om_p$).
The Alfv\'{e}n-O mode satisfies the 
 dispersion relation
\be
n_{A-O}^2 = \frac{   (\om^2 -  2 \om_p^2)(\om^2 -  2 \om_p^2 - \om_B^2) 
}{ (\om^2 -  2 \om_p^2) (\om^2 - \om_B^2 ) - 2 \om_B^2 \om_p^2 \sin^2 \theta }
\label{2} \ee
 Alfv\'{e}n branch is always subluminal  while O mode
is {\it superluminal} at small wave vectors and 
{\it subluminal} at large wave vectors.

In the limit $\om_p\ll \om$ we find
\be
(\Delta n) = n_X-n_{A-O} = - \frac{\om_B ^2 \om_p^2}{\om^2 (\om^2 -\om_B^2)} \sin^2\theta 
 =  
\left\{
\begin{array}{cc}
- \frac{ \om_B ^2 \om_p^2}{\om^4} \sin^2\theta, & \om_B \ll \om
\\
\frac{  \om_p^2}{\om^2} \sin^2\theta, & \om_B \gg \om,
\end{array} 
\right.
\label{deltan}
\ee
The $\om_B \ll \om$ has been discussed previously \citep{1969SvA....13..396S,1997PhRvE..56.3527M,1998PASA...15..211K}. The $\om_B \gg \om$ is the new regime of interest:   in highly magnetized plasma the phase  velocity difference is independent of the \Bf\ (the X-mode is nearly luminal).

\subsection{Faraday Conversion  in magnetized  pair plasma}

\begin{figure}[h!]
\includegraphics[width=0.99\linewidth]{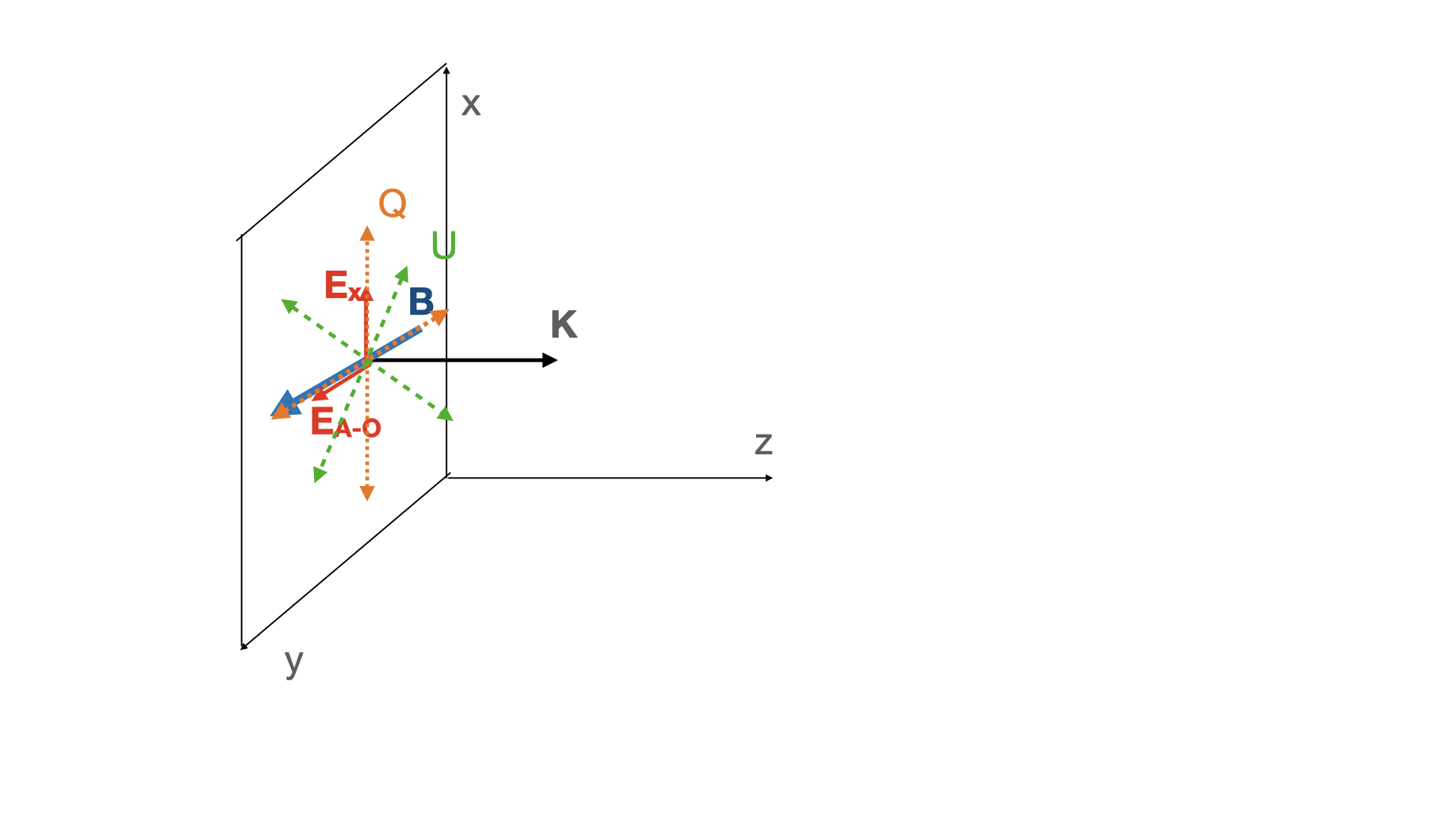}
\caption{Coordinates, normal modes and Stokes parameters. The wave propagates along $z$ direction, \Bf\ is along $y$ direction, X-mode is polarized along $x$, A-O-mode is   polarized along $y$, Stokes' Q corresponds to pure oscillation  of \Ef\ either along $x$ or $y$ (the $+$), Stokes' U corresponds to pure oscillation  along the $\times$ axes.  In such coordinates the  Stokes' Q represents a normal mode, hence there is no Faraday conversion. Stokes U  is a mix of two a normal modes with different phase velocities.}
\label{coords}
\end {figure}

Consider propagation of \EM\ radiation along $z$ direction perpendicular to the \Bf, which is in the $y$ direction, Fig. \ref{coords}

The X-mode is then polarized along $x$ and {A-O}-mode along $y$.
Let at the point $z=0$ the wave be  linearly  polarized with  the unit  Jones  \citep{1980poet.book.....B} vector 
\be
\E_0 = 
\left(
\begin{array} {c} 
\cos \chi_0
\\
 \sin \chi_0
 \end{array}
 \right)
\ee

The medium works as a retarder. 
The key parameter is a phase lag (retardance) $\delta$:
\be
\frac{d\delta}{dz}=   \frac{\om}{ c} (\Delta n) 
\label{ddelta0}
\ee

In the limit $\om_B \gg \om_p, \om$,
\ba  &&
\frac{d\delta}{dz}=  \frac{  \om_p^2}{\om c} = \frac{2 e^2 n}{c^2 m_e} \times \lambda= \delta_0 \times \lambda
\nn &&
\delta_0 = \frac{2 e^2 n}{c^2 m_e} 
\nn && 
\delta = \int \frac{d\delta}{dz} dz=  k \int (\Delta n)  dz
=
 \lambda  \frac{ 2 e^2}{m_e c^2} n \Delta z = 2 r_e \lambda ( n \Delta z)
= 1.7 \times 10^6 \lambda \times  {\rm DM}
\label{ddelta} 
\ea
where $ (\Delta n) $ is the difference of the refractive indices of the two  linearly polarized modes,  with $\theta = \pi/2$ assumed; the limit "$\to$" here and below corresponds to the relevant case of $\om_B \gg \om$; dimension of $\delta_0$ is cm$^{-2}$. In the last relation  $\lambda $  is in centimeters and DM is the dispersion measure in pc  cm$^{-3}$.

It may be convenient to define the magnetic Conversion Measure (CM$_{\rm B}$) as
\ba &&
\delta = {\rm CM}  _{\rm B} \times \lambda, \, \om_B \gg \om
\nn && 
 {\rm CM}_{\rm B} = \frac{ 2 e^2}{m_e c^2} n \Delta z  = 1.7 \times 10^6  \times  {\rm DM}\, {\rm cm}^{-1}= 1.7 \times 10^4  \times  {\rm DM}\, {\rm m}^{-1}
 \label{CM}
 \ea
 It  is different from the  low \Bf\   Conversion Measure (CM)
\ba &&
\delta = {\rm CM}  \times \lambda^3 , \, \om_B \ll \om
\nn && 
 {\rm CM} = \frac{  e^4}{2 \pi^2 m_e^3  c^6} n \Delta z  B^2 = 1.4 \times 10^{-2}   \times  {\rm DM}  \times B^2 \, {\rm cm}^{-3}=  1.4 \times 10^{-8}   \times  {\rm DM}  \times B^2 \, {\rm m}^{-3}
 \label{CM1}
 \ea
 where \Bf\ is in Gauss. 

Relations (\ref{ddelta}-\ref{CM})  demonstrates that  in magnetically-dominated plasma a screen with DM $\sim 10^{-6}$  pc cm$^{-3}$ can induce polarization changes (rotation of PA of linear component, as well as production of circular component) of the order of unity. This is one of the major points of the work.

Using Jones' calculus, at any location the polarization can be characterized by a vector
\ba &&
\E= 
\left(
\begin{array} {c} 
\cos \chi_0
\\
 e^{i \delta}  \sin \chi_0
 \end{array}
 \right) 
 = \hat{J} \cdot \E_0
 \nn &&
 \hat{J}= \left(
\begin{array} {cc} 
1 &0 
\\
 0 & e^{i \delta}  
 \end{array}
 \right) 
\ea
$\hat{J}$ is Jones' matrix.

The corresponding Stokes parameters (normalized to unity; fully polarized wave is assumed) are
\be
\hat{P} = 
\left(
\begin{array} {c} 
Q
\\
U
\\
V 
\end{array}
 \right) =
 \left(
\begin{array} {c} 
\cos (2 \chi_0) 
\\
\cos \delta \sin (2 \chi_0)  
\\
\sin \delta \sin (2 \chi_0)  
\end{array}
 \right) =
 \left(
\begin{array} {c} 
Q_0 
\\
 U_0 \cos \delta
\\
 U_0\sin \delta
\end{array}
 \right) 
 \label{P} 
\ee

Polarization transfer equation can be written as
\ba && 
\partial_z \hat{P} = {\bf \Omega} \times  \hat{P} 
\nn &&
 {\bf \Omega} = \{ \partial _z \delta, 0,0 \} \to   \{ 2 \frac{  \om_p^2}{\om c} , 0,0 \} 
 \label{Omega}
 \ea
 ${\bf \Omega} $ is the angular frequency of the polarization rotation rate on the Poincare sphere.

\begin{figure}[h!]
\includegraphics[width=0.99\linewidth]{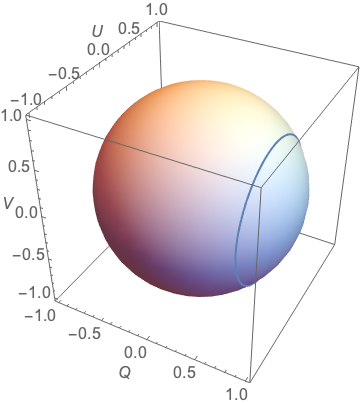}
\caption{Example of trajectory on the Poincare sphere for initial polarization in the   $\chi_0 =  \pi/8$ state ($Q=U=1/\sqrt{2}$). Parameter $Q$ remains constant.}
\label{UQV}
\end {figure}

The corresponding Mueller matrix
\be
{\bf M}= 
\left(
\begin{array}{cccc}
 1 & 0 & 0 & 0 \\
 0 & 1 & 0 & 0 \\
 0 & 0 & \cos  \delta & -\sin  \delta \\
 0 & 0 & \sin  \delta & \cos  \delta \\
\end{array}
\right)
\ee
Thus, there are periodic $U-V$ oscillations. (It can be called $\sim$ generalized Faraday conversion, but the proper Faraday effect - as opposed to Cotton-Mouton effect -   is not involved in this case). 

\begin{figure}[h!]
\includegraphics[width=0.99\linewidth]{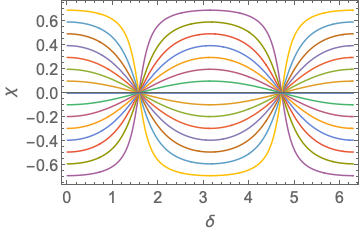}
\caption{Position angle $\chi$ of  linear  polarization as function of retardance $\delta$ for $\chi_0$ between $0$ and $\pi/2$ with a step $\pi/32$ ($\chi_0=\pi/2$  is omitted - this would give a straight line at $\chi =   \pi/2$). Values of $\chi=0$ correspond to maximal circular  polarization $V=U_0$ and  $U=0$. }
\label{chiofdelta}
\end {figure}

The electric field traces an ellipse with ellipticity
\be
\epsilon = 
\frac{\sqrt{\sqrt{1+\cos ^2 \delta \tan ^2\left(2 \chi_0\right)}- \cos ^2\delta 
   \sin \left(2 \chi_0\right)\tan \left(2 \chi_0\right)+\sin
   ^2\left(\chi_0\right)-\cos ^2\left(\chi_0\right)}}{\sqrt{\sqrt{1+ \cos ^2\delta 
   \tan ^2\left(2 \chi_0\right)}+\cos ^2 \delta \sin \left(2 \chi_0\right) \tan
   \left(2 \chi_0\right)-\cos \left(2 \chi_0\right)}}
   \ee
   and position angle 
   \be
   \tan (2 \chi)= \cos \delta \tan (2 \chi_0),
   \label{chi}
   \ee
   see Fig. \ref{chiofdelta}.
   
Thus,   in  symmetric pair plasma there is efficient transformation  of Stokes parameters $U$ and $V$, in a way similar to the   laboratory  phase retarders. The key parameter for the rate  is the  retardance $\delta$, Eq.  (\ref{ddelta}).  The rotation of polarization/production of the circular  component  in pair plasma disappears only  for $U=0$ with $2 \chi_0 = 0, \, \pi$; this  corresponds to either pure X mode ($ \chi_0 = 0$) or pure A-O mode ($ \chi_0 =\pi/2$).

\subsection{Frequency scaling of Generalized Faraday Rotation}

In plasma the retardance  (\ref{ddelta}) and the rotation angle (\ref{chi}) are wave length-dependent. The polarization angle scales as 
\be
\partial _\lambda \chi=
\frac{\sin (\delta ) \tan \left(2 \chi _0\right)}{2 \left(1+\cos ^2(\delta ) \tan
   ^2\left(2 \chi _0\right)\right)}
   \partial _\lambda \delta
   \label{RM}
   \ee
Relation (\ref{RM}) can be interpreted as a Generalized Faraday Rotation (GFR): frequency-dependent PA. Note that when the retardance  crosses $\delta = \pi$ the sign of $\partial _\lambda \chi$, and the corresponding RM,  changes \citep[][]{2022arXiv220308151D}

   At small $\delta \ll 1$ the frequency dependent part of the PA scales as
   \be
  \chi  =  \frac{ \sin ( 4 \chi_0)}{8}  \left(  \int \delta_0 dz \right)^2   \times \lambda^2
  \ee
  Thus,  frequency scaling matches the Faraday effect. But the polarization rotation in this limit is  {\it independent} of the \Bf\ (and hence cannot be used to estimate it).
  We can introduce effective RM,  $ {\rm RM}_{eff} $,
  \ba && 
 \chi    =   {\rm RM}_{eff}  \times  \lambda^2
  \nn &&
    {\rm RM}_{eff} = {10^{-4}}  \times \frac{ \sin ( 4 \chi_0) }{8} \left(  \int \delta_0 dz \right)^2=
    \nn &&
     {10^{-4}}   \times  \sin ( 4 \chi_0) \frac{e^4}{ 2 m_e^2 c^4} n ^2 ( \Delta z)^2=
3 \times 10^{11}   \times \sin ( 4 \chi_0)  \times  {\rm DM}^2 \, {\rm m}^{-2}
    \label{RMeff}
    \ea
    (factor of $10^{-4}$ converts from cgs unites  to m$^{-2}$). 
        Thus, to produce RM = $10^5$ in 
    FRB121102 \citep{2018Natur.553..182M} the required DM is only  $5 \times 10^{-6}$.

    At special moments when $ \delta \approx \pi/2$ we find from (\ref{chi})
    \be
     \Delta \chi  = \tan{2 \chi_0}  \Delta[ \delta] = 
     \frac{ e^2 n z \tan \left(2 \chi _0\right)}{ m_e c^2} \times  \Delta[\lambda] =8 \times 10^5 \tan (2 \chi _0)   {\rm DM}  \times  \Delta[\lambda]
     \ee
  At this point circular component is large.
   Note that in both cases  $\delta  \ll 1$ and  $\delta  \sim \pi/2$ the rotation angle is {\it independent} of the \Bf. 
   
  Generally we can write
  \ba && 
    \Delta \chi \propto  \Delta[\lambda^p]
    \nn &&
    1<p< 2
    \ea
    In astrophysical literature this is usually called Generalized Faraday Rotation.

In conclusion, we arrived at two important results: (i) there is efficient production of circular component in symmetric pair plasma; (ii) the corresponding  frequency scaling (in the $\delta \ll 1$ regime) matches the Faraday  rotation.   Thus, for the observed radio signal, if  $\chi_{\rm high}  -\chi_{\rm low}$, which is the rotation angle difference between the top and bottom frequency, is much smaller than 1 radian, the Generalized Faraday rotation may be confused with Faraday rotation and misinterpreted as large RM or sign change of RM (although in the case of FRB 20190520B, the observed change of PA against frequency is large enough to distinguish different effects.)

In what follows we apply the general relations derived above to relativistically streaming plasma in the inner parts of the pulsar/magnetar winds.
 
 \subsection{Faraday Conversion in electron-ion  plasma}
 
 All the above relations, when expressed in terms of retardance $\delta$, Eq.  (\ref{ddelta0}), are  applicable to regular electrons-ion plasma for propagation orthogonally to the \Bf,  when the normal modes are linearly polarized. In particular, 
 a term similar to the $\om_B \ll \om$ limit in (\ref{deltan}) appears also in non-symmetric plasma. In that case for wave  propagating  orthogonal to the \Bf, 
\be
(\Delta n)_{e-i}= - \frac{ \om_B ^2 \om_p^2}{2 \om^4} 
\label{dnnorm} 
\ee

 The principles of Faraday conversion remain the same: in a frame  defined in Fig. \ref{coords}, the Stokes Q remains constant while U  and V experience oscillations.
 The rotation direction on the Poincare sphere $ {\bf \Omega}$, Eq. (\ref{Omega}),  is still aligned with Q.

  In regular plasma at small $\om_B,  \, \om_p \ll \om$ the retardance is, see Eqns.  (\ref{dnnorm}) and (\ref{ddelta0}),
 \be
 \frac{d\delta _{e-i}}{dz}= -  \frac{ \om_B ^2 \om_p^2}{2 c  \om^3} 
 \label{deltaei}
 \ee
 It has different sign and different frequency scaling than the $\om_B \gg \om$ case.  The motion  on the Poincare sphere in this case proceeds in the opposite sense around Q axis (counter-clockwise instead of clockwise). 
 
 The sense of rotation on the Poincare sphere  cannot be used to distinguish the two cases  observationally:  the Q-U separation is observer-dependent (typically Q is  chosen along  the direction to the north). There is then a freedom in rotation on the Poincare sphere around V axis. 
Qualitatively:  the difference between the phase speeds  of the   X-mode and   the A-O-mode changes sign at the resonance, but  observationally we do not know the direction of the \Bf:  hence cannot define which mode is which: a change of \Bf\ direction by $\pi/2$ ``flips''  the  observational definition of the X and   A-O-modes. 
 
In constant density/\Bf\ the total retardance (\ref{deltaei}) evaluates to
\be
\delta _{e-i} = \frac{2 \pi e^4}{m_e^3 c^3 \om^3} \times {\rm DM} \times B^2 = 7 \times 10^{-3} \lambda^{3} {\rm DM}\, B^2
\label{deltaei1} 
\ee
where DM is the dispersion measure in pc cm$^{-3}$, $\lambda$ is in cm,  and \Bf\ is Gauss. The condition $\delta _{e-i}  \sim 1$ is much more demanding than (\ref{ddelta}).

\section{Magnetar/pulsar  winds}

\subsection{Particle dynamics in the  inner  wind, $R_{LC} \leq r \leq  r_0$}
\label{bead}

Let us consider motion of particles in the inner wind with  \cite{1973ApJ...180L.133M}  \Bf\ and  assuming that particles moves exclusively along the field (``bead-on-wire" approximation"). This is best done using machinery  of General Relativity \citep[][]{LLII}.

In spherical coordinates, 
changing to the rotating  system of coordinates
$ d\phi \to d \phi' - \Omega dt$, 
and assuming that particles move along the Archimedean spiral with 
\be
d\phi' = - \Omega dr
\ee
at fixed   polar angle $d\theta_p=0$, we
 find the metric tensor
 \ba && 
 g_{00} = - (1-r^2 \sin^2 \theta_p \Omega^2)= - g^{rr}
 \nn &&
 g_{rr} = (1+r^2  \sin^2 \theta_p \Omega^2) = - g^{00}
 \nn &&
 g_{rt} = r^2  \sin^2 \theta_p \Omega^2= g^{rt} 
\ea

The Hamilton-Jacobi equation
\be
g^{\alpha \beta} \partial_\alpha  S  \partial_\beta  S =1
\ee
for  the action functional $S$,
with separation 
$
S = - \gamma_0 t + S_r (r)
$
becomes
\be 
\gamma _0^2 \left(r^2 \Omega ^2 \sin ^2\theta_p+1\right)-2 \gamma _0 r^2 \Omega ^2 \sin
   ^2\theta_p S_r'(r)-S_r'(r){}^2 \left(1-r^2 \Omega ^2 \sin ^2\theta_p\right)=1
   \label{HJ1}
   \ee
   The \Lf\ $\gamma_0$ comes  mostly from  the motion  of particles {\it along} the \Bf.

   This can be integrated to find $S_r$. Differentiating  the result with respect to $\gamma_0$ we find
   \be 
  t = r  - \frac{1}{ \sin \theta_p \Omega}  + \ln \left( \left(1-\frac{1}{\gamma _0^2}\right)  \frac{(1-r \Omega  \sin \theta_p)
   \left(\sqrt{\gamma _0^2+r^2 \Omega ^2 \sin ^2\theta_p-1}+\gamma _0 r \Omega  \sin
   \theta_p\right)}{(1+ r \Omega  \sin \theta_p) \left(\sqrt{\gamma _0^2+r^2 \Omega ^2
   \sin ^2\theta_p-1}-\gamma _0 r \Omega  \sin \theta_p\right)}\right) \frac{1}{2  \sin \theta_p \Omega} 
       \label{HJ2}
   \ee
Eq.    (\ref{HJ2}) gives implicit  solution for motion of a bead along the Michel-type \Bf\ lines; time is chosen $t=0$ at the moment when the particle crosses the \LC, $r = 1/(\sin \theta_p \Omega)$.
  
   By differentiating with  respect to time we find  coordinate velocity
   \ba && 
 \beta_r \equiv   \partial_t r = 
  \frac{ \varpi^2 \Omega ^2}{1+  \varpi^2 \Omega ^2} \left( 1-\frac{1}{\gamma _0^2+\gamma _0^2  \varpi^2 \Omega ^2+ \varpi^4 \Omega ^4 }\right) 
   +
    \frac{\gamma _0 \sqrt{\gamma _0^2+ \varpi^2 \Omega ^2 
   -1}}{\gamma _0^2+\gamma _0^2  \varpi^2 \Omega ^2 + \varpi^4 \Omega ^4 } 
   \nn &&
   \approx
   \left\{
   \begin{array}{cc}
    1 - \frac{1}{ \gamma_0^2}, &  \Omega  \varpi \ll  \gamma_0
    \\ 
    1 - \frac{1}{  \varpi^2 \Omega ^2},  & \, \mbox{for} \,   \varpi \to \infty
    \end {array}
    \right. 
    \nn &&
    \varpi = r \sin \theta_p
   \label{Gamma2}
   \ea
   
   The toroidal component of the velocity
   \be
   v_\phi=  \varpi  (1-\beta_r) 
    \label{Gamma3}
   \ee
   remains small; it's maximal value is reached at $\approx 1.27 \gamma_0 / \Omega$ and equals $v_{\phi, max} \approx 0.3/\gamma_0$ (at $\theta_p =\pi/2$). Particles move nearly radially.

Relation (\ref{Gamma2}-\ref{Gamma3}) requires some explanation. Motion of particles  consists of (i) bulk E-cross-B drift  and (ii)  motion  along the field.  The bulk E-cross-B drift has two components:  radial and toroidal   \citep{Michel73}
\be
{\bf v} = \{ \frac{  \varpi^2}{1+ \varpi^2} ,0,\frac{  \varpi}{1+ \varpi^2}\}
\ee
In the inner part of the wind the unit \Bf\ vector
\be
{\bf e}_B =  \{ \frac{  1}{\sqrt{1+ \varpi^2}} ,0, - \frac{  \varpi}{\sqrt{1+ \varpi^2}}\}
\ee
quickly becomes toroidal (the field lines intersect the \LC\ at $45^\circ$). As a result, near the  \LC\ the large azimuthal drift velocity is mostly compensated by the azimuthal component of the parallel velocity, resulting in nearly radial motion with relativistic \Lf\ $\gamma$.  Particles stream  radially  with \Lf\ $\gamma_0$ (injection \Lf). The EM waves propagate across magnetic field, $\theta= \pi/2$. 

In conclusion, in the inner part of the wind, somewhat outside the \LC\ particles move nearly radially with the \Lf\ $\gamma_0$ determined by the acceleration processes {\it inside} the \ms.

\begin{figure}[h!]
\includegraphics[width=0.99\linewidth]{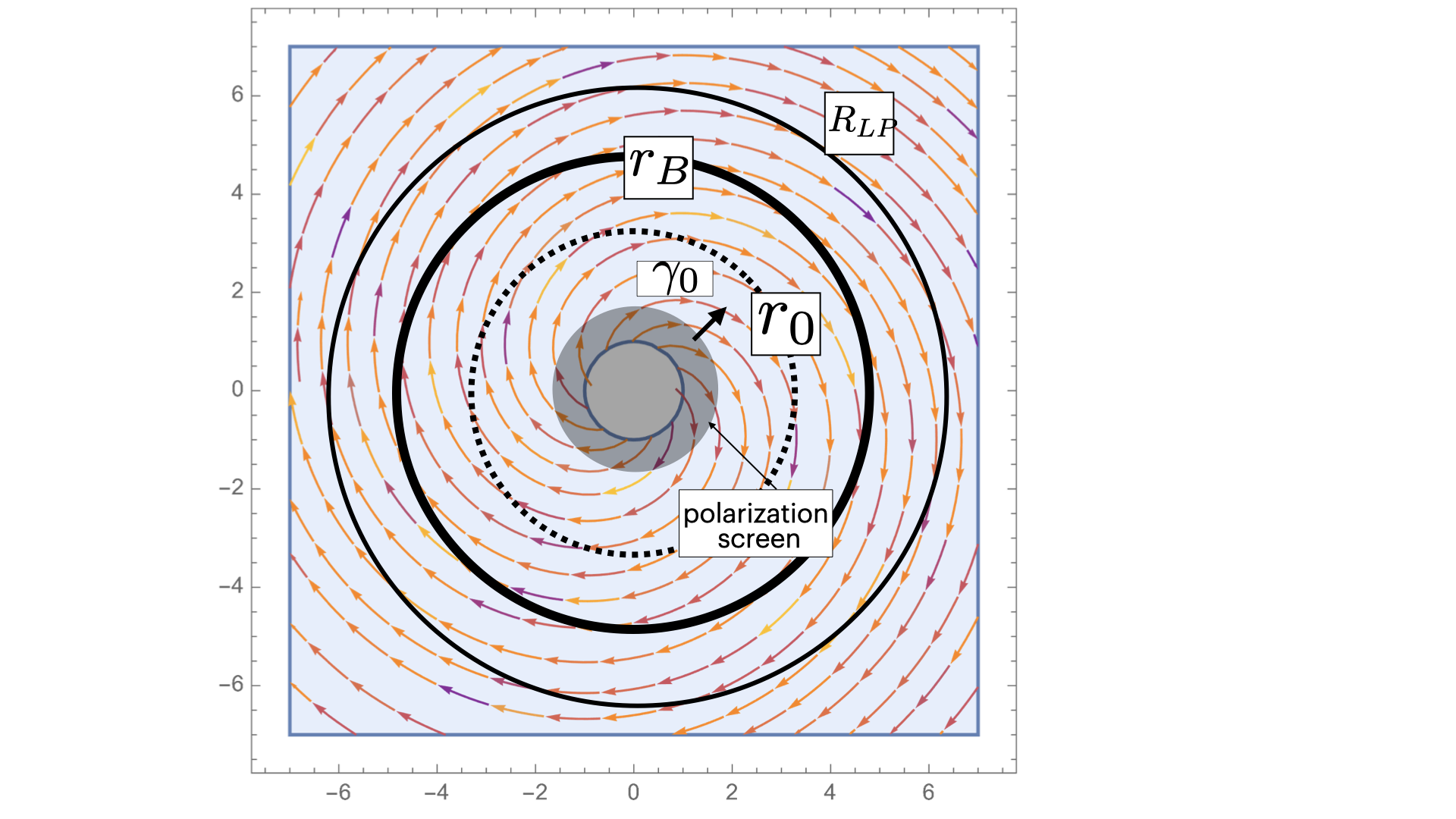}
\caption{ Geometry of PA  rotation in the inner parts of the wind (not to scale). Arrows are \Bf\  \protect\citep{1973ApJ...180L.133M}. Particles leave the \ms\ with \Lf\ $\gamma_0$, moving nearly radially. At distance $r_0 \sim {\gamma_0} R_{LC}$  the \Lf\ starts to increase $\gamma \propto r/R_{LC}$.  Cyclotron resonance occurs at $r_B$, limiting polarization radius is  $R_{LP}$. Depending on parameters of the flow relative location of $r_0$, $r_B$ and $R_{LP}$ may change.}
\label{Faraday-pair-pic}
\end {figure}

\subsection{Wind parameters}

Let us parametrize the properties of the wind by wind luminosity $L_w$ and the ratio of Poynting to particle fluxes $\mu$:
\ba &&
\mu =\frac{L_w}{ \dot{N}  m_e c^2} =  \frac{B_{LC}^2}{4\pi n_{LC} m_e c^2} 
\nn &&
B_{LC}= \frac{\sqrt{L_w} \Omega}{c^{3/2}}
\nn &&
 n_{LC} = \frac{L_w}{4\pi \mu m_e c^3 R_{LC}^2 }
\label{7} \ea

Terminal Lorentz factor of the wind is $\Gamma_w$;  it is  reached at $r_w$, 
\ba &&
\Gamma_w = \mu^{1/3}
\nn &&
r_w =  \mu^{1/3} R_{LC}
\label{rw}
\ea
\citep{1969ApJ...158..727M,1970ApJ...160..971G,1973ApJ...180L.133M}. Requirement of $\Gamma_w \geq \gamma_0$ then gives $\mu \geq \gamma_0^3$. This is a requirement that the terminal \Lf\ is determined by the wind acceleration, not by the injection.

If density is scaled to \cite{GJ} density $n= \kappa n_{GJ}$, then
\be
\kappa \mu = \frac{e \sqrt{L_w} }{m_e c^{5/2}} = 3 \times 10^{10} L_{w, 38}
\label{kappa}
\ee


\subsection{Cyclotron resonance}

One of the key issues is the location  the  cyclotron resonance $\om' \sim \om_B'$. For a given frequency $\om$ (in the observer frame)  the cyclotron resonance occurs at
\ba && 
r_B = \frac{e \sqrt{L_w}}{c^{3/2} m_e \om}
\nn && 
\frac{r_B }{R_{LC}} = 30 L_{w, 38} \nu_9^{-1} P^{-1}
\label{9} \ea
where period is in seconds. (Note that for Crab pulsar, $P =0.033$ seconds,  the cyclotron resonance occurs at $\sim 10^3 R_{LC}$.) Relation (\ref{9}) is independent of the \Lf\ of the wind.

\subsection{ Limiting polarization radius}
\label{limiting}
Separation of modes into X and O branches may be
violated if the rate of change of plasma parameters is sufficiently fast, so that the mode propagation becomes non-adiabatic \cite[the effect of limiting polarization][]{1952RSPSA.215..215B}. This occurs when the wavelength of the beat between two modes becomes larger than the scale at which the properties of the modes change.  In our case this condition becomes
\be
\left( \frac{\om}{\Gamma c} \right) \left( \frac{r}{\Gamma } \right) \left( \Delta n\right)^\prime \geq 1
\label{23} \ee
where $\Gamma$ is the \Lf\ of the wind, which follows from (\ref{Gamma2}).
This is a condition that propagation is adiabatic.

The condition for limiting polarization becomes
\be
R_{LP} = 2 \frac{e^2}{m_e^2 c^4} \frac{L_w}{\Gamma \mu \om} =
  \left\{
\begin{array}{cc}
 2 \frac{e^2}{m_e^2 c^4} \frac{L_w}{\gamma_0 \mu \om}  , & R_{LP} \leq r_0
 \\
  \frac{\sqrt{2} e \sqrt{L_w} \sqrt{R_{LC}}  }{m_e c^2  \sqrt{\mu} \sqrt{\om}  } , & R_{LP} \geq r_0
 \end{array}
\right.
\ee
Location of the limiting polarization radius is thus highly dependent on the parameter $\gamma_0 \mu$. For definiteness as assume 
 $R_{LP} \geq r_0$.

\section{Polarization transfer in magnetar winds}

\subsection{Retardance and  Generalized Rotation Measure in the wind}

\begin{figure}[h!]
\includegraphics[width=0.9\linewidth]{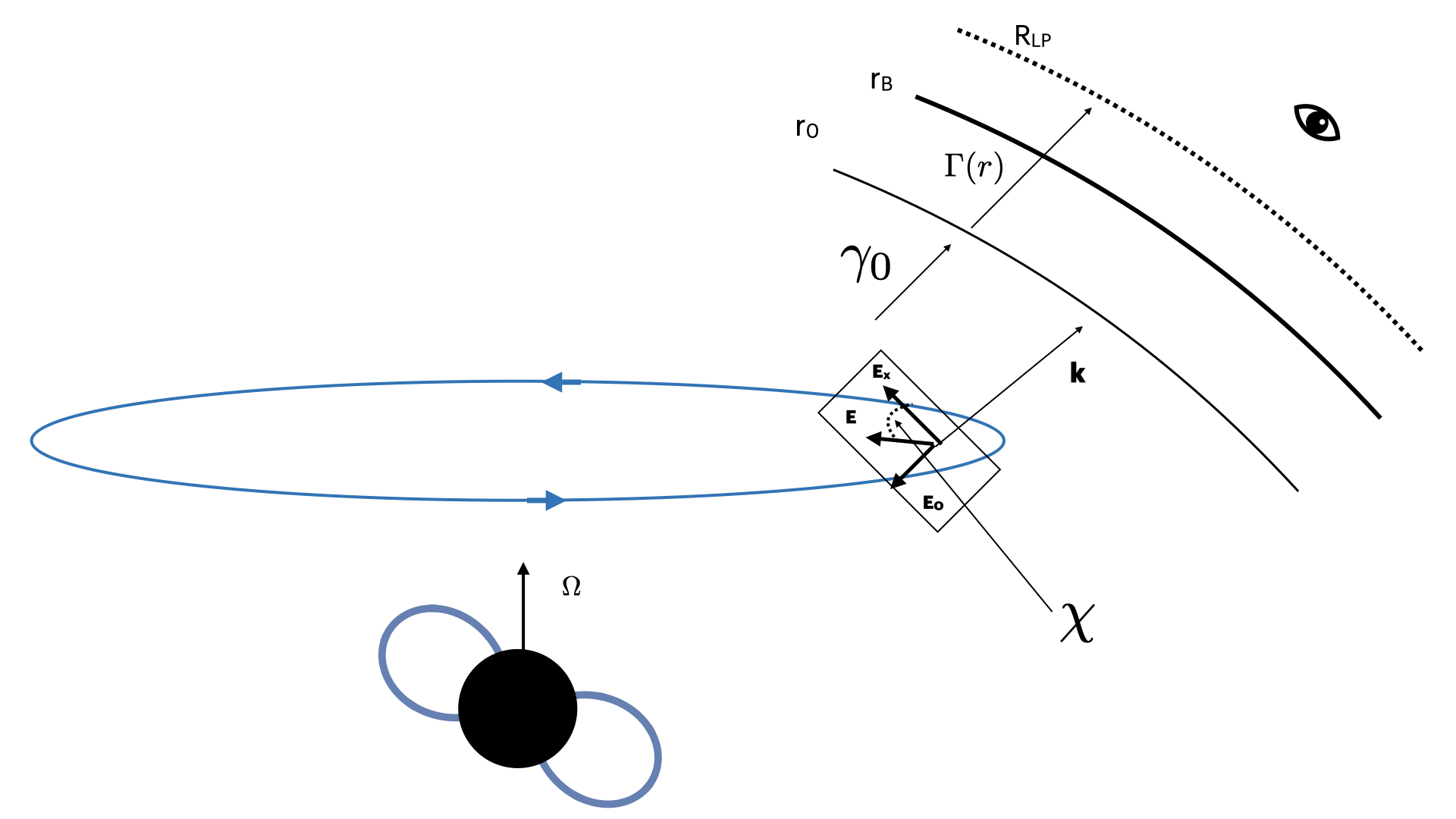}
\caption{Particle dynamics and polarization in the inner wind. The \Bf\ quickly becomes nearly toroidal, particles stream nearly radially.  Until $r\sim r_0 =\gamma_0$ the \Lf\ is the injection \Lf, determined by magnetospheric processes, at larger radii the wind accelerates linearly. Radiation produced in the \ms\ can be separated into X and O modes. The linear component of polarization makes angle $\chi$ with respect to the projection of the spin of the \NS\ on the sky. }
\label{Polarization-picture001}
\end {figure}

Let us assume the following scaling in magnetar winds: 
$r_{LC} \leq r_0 \leq   R_{LP}, \, r_w $, see Fig. \ref{Polarization-picture001}, so that the cyclotron radius is outside the constant \Lf\ region $r_0$. 
Let us consider polarization transformation in  the  regions $r < r_0 < r_B$, where $\om_B \gg \om$ and the \Lf\ is constant $\sim \gamma_0$.

As we demonstrated in \S  \ref{bead} azimuthal motion of particles 
 can be neglected.   The wind can  then be approximated  as a sequences of toroidal
magnetic loops moving  away from the light cylinder. 
The EM waves propagate across magnetic field, $\theta= \pi/2$ through relativistically moving  wind.

In this regime the \Lf\ of the particles is $\sim \gamma_0$, and $\om \leq \om_B$. 
The rate of retardance in the wind frame 
\be
\left( \frac{ d \delta}{ d t'}  \right)^\prime =  \frac{\om_p^{\prime, 2} }{\om'}
\ee
Transformation to lab frame gives
\be
\frac{ d \delta}{ d t} =  \frac{\om_p^{ 2} }{ \gamma_0 \om}
\ee
(values of $\om_p^2$ and $\om$ transform similarly, while the rate is smaller in the observer frame.)

Estimating $\Delta z \sim R_{LC}$, density according to (\ref{7}), we find 
\be
\delta \sim \frac{e^2}{2\pi m_e^2 c^6} \frac{L_w \Omega \lambda}{\mu \gamma_0}  = 3.8 \times 10^{10} \, \frac{L_{sd, 38} \lambda}{ P  \mu \gamma_0}
\label{delat10} 
\ee

The key dependence is on the combination of  parameters $(\mu \gamma_0)$. It is very large, a product of two large numbers. For example,  for Crab pulsar with multiplicity  $\kappa \sim 10^4$, this  gives $\mu \sim 4 \times 10^6$; for typical $\gamma_0 \sim $ few $\times 10^3$,  we find $ \mu \gamma_0 \sim   10^{10}$. 

To get  large phase shift $\delta \geq  1$ requires
\be
\mu \gamma_0  \leq  \frac{e^2}{2 \pi m_e^2 c^6} {L_w \Omega \lambda}=
3.8  \times 10^{11} \, L_{w,38}  \lambda P^{-1}
\ee
(period $P$  in seconds). 

The effective RM (\ref{RMeff}) then becomes 
\be
   {\rm RM}_{eff}  =3 \times 10^{-7}   \frac{ \sin (4 \chi_0)}{  (\mu \gamma_0)^2 } \frac{e^4}{m_e^4 c^{12}}{L_w^2 \Omega^2}
     \label{RM1}
\ee
(this assumes $\delta \leq 1$).

To produce an observed RM of value $RM_{ob}$, the ratio of Poynting to particle fluxes and the streaming \Lf\ should satisfy the condition 
\be
\mu \gamma_0   \leq 1.4  \times  10^{8}  \times  \sqrt{ \sin (4 \chi_0)}  \frac{ L_{w,38} }{P\,   \sqrt{ {\rm RM_{ob}}} }
\ee
(smaller $\mu$ implies larger observer frame density,  smaller $\gamma_0$ implies less time dilation).

In particular for $RM_{ob}=10^5$  inferred by   \cite{2018Natur.553..182M} it is required that
\be
\mu \gamma_0  \leq 5 \times  10^{5}  \times  \sqrt{ \sin (4 \chi_0)}  \times  \frac{ L_{w,38} }{P}
\label{mugamma0}
\ee
Relations (\ref{delat10}-\ref{RM1}) are our main results. They provide estimates of the retardance and the Rotation Measure through the inner part of the wind (expression for the RM$_{eff} $ assumes $\delta \leq 1$).

 Models of pair creation in the pulsar \mss\ \citep{1975ApJ...196...51R,1979ApJ...231..854A,2010MNRAS.408.2092T} typically predict $\gamma_0 \sim 10^3-10^5$ and similarly for $\kappa  \sim 10^3-10^5$. Thus,   for Crab pulsar $\mu \sim 10^5-10^7$ and $\mu \gamma_0 \sim 10^9-10^{11}$, Eq. (\ref{kappa})).  Such  plasmas  are too rarefied and too fast to produce observable effect,  Eq. (\ref{RM1}).
 
 Magnetar plasmas are expected to be much denser and slower. First, the basic current density can be much higher than the Goldreich-Julian \citep{tlk}. Second,  the bulk \Lf\ are expected to be  smaller, $\gamma_0 \sim 10^2$  \citep{2013ApJ...777..114B}.  For this value of the \Lf\ the Eq. (\ref{rw}) then implies that in order to satisfy (\ref{mugamma0}) the wind must be heavy loaded with  $\mu \leq  10^3$; thus $\mu \leq \gamma_0^3$ - so that the terminal \Lf\ of the wind is determined by the injection \Lf\ $\gamma_0$, not acceleration  of the wind.

\subsection{Polarization evolution  near the  cyclotron resonance}

The region near the cyclotron resonance $ r_B$ presents a challenge, both in terms of the possibility of cyclotron absorption,  large rates of PA rotation, and harder to quantify effects of the \Lf\ spread. If cyclotron absorption is negligible, \S \ref{Cyclotron}, for mono-energetic beam,  large rotation angle near the resonance will be mostly cancelled, since at two sides of the resonance the rotation direction is in the opposite sense. But a small mismatch between inner and outer parts may produce large net rotation near the resonance.

Near the cyclotron  resonance the evolution of the retardance rate $ d \delta/dz$  becomes infinitely fast (for mono-energetic beam - to be smoothed out due to velocity spread). Most importantly, $ d \delta/dz$ changes sign as the wave goes through the resonance.  If the plasma density at both sides of the resonances  were the same then (in the absence of absorption) the total  PA rotation angle would be zero. In the expanding  wind the plasma density``after'' the resonance is smaller than ``before. This creates finite PA rotation
 as the EM pulse propagates through the resonance.  
 
 Integrating (\ref{ddelta})  with  (\ref{deltan})  we find for the resonant contribution
 \be
 \left. \delta \right|_{\rm res} \approx   \frac{ R_{LC} \om_{p, r_B} ^2}{2  c \om}   \ln \left(  \frac {r_B}{R_{LC}}  \right) \approx \frac{R_{LC} \om}{ 4 c \mu} \ln (r_B/R_{LC} ) \approx
 5 \times 10^3 \mu_6^{-1} P \lambda^{-1}
 \label{chiLC1}
 \ee
 where $\om_{p, r_B} $ is the plasma frequency at the resonance. Qualitatively, retardance   (\ref{chiLC1}) resembles (\ref{ddelta}), with two modifications: (i)  it's the density at the cyclotron resonance that appears in  (\ref{chiLC1}) (this reduces the rotation angle); (ii)  the logarithm $\ln (r_B/R_{LC} ) \sim 10 $ appears due to density disbalance at two sides of the resonance.
Total PA rotation through resonance can be large. 
 
 Notice that retardance  through the resonance is {\it inversely } proportional to wavelength: for longer waves the location of the cyclotron  resonance is proportional to $\lambda$, so that  density at the resonance $\propto \lambda ^{-2}$. This reduces the scaling $\delta \propto  \lambda$ in constant density/\Bf\  (\ref{ddelta}) to $\delta \propto  \lambda^{-1} $
in  decreasing  density/\Bf.

\section{Cyclotron absorption in the wind}
\label{Cyclotron} 

\subsection{General relations}

The resonant  optical depth can be  estimated as \citep[][]{1996ASSL..204.....Z,1994ApJ...422..304T,lg06}
\ba && 
\tau_{res} \approx \sigma_{res}   n'\frac{r}{\Gamma}
\nn &&
\sigma_{res} = \frac{\pi^2 e^2  }{m_e c \om_B'}
\nn &&
\tau_{res} =\frac{ e \sqrt{L_w}}{ m_e c^{5/2} \mu \gamma_0} = 2 \times  10^{10} \frac{ L_{w,38}^{1/2} }{\mu \gamma_0}
\label{15}
 \ea
Given the limit on the product  $(\mu \gamma_0)$ from above, Eq. (\ref{mugamma0}), optical depth to cyclotron absorption is large.

In strong \Bf\ a particle quickly emits a cyclotron-absorbed photon:  in this case the process is resonant scattering, not absorption. 
Cyclotron emission time in the frame of the wind is
\be 
t_c ' \approx  \frac{ m_e^3 c^5}{  B^{\prime , 2} e^4}
\ee
Comparing with flight time in  the rest frame $(r/\gamma_0)/c$
\be
\frac{ c  t_c '}{r/\Gamma} =\frac{ m_e^2 c^{11/2}   \gamma_0^3}{   e^3  \sqrt{L_w} \om}= 5 \times 10^2 \, \gamma_0^3 L_{w,38}^{-1/2} \gg 1
\ee
Thus,  it is absorption, not scattering.

Absorption will affect mostly the X-mode (for propagation perpendicular to the \Bf\ the A-O mode is not affected). Fluctuations of wind density may still allow escape of the X-mode, and of the circularly polarized component.  

\subsection{Cyclotron absorption in Crab pulsar}
\label{Crab} 

Crab pulsar presents an interesting case: using the fact that we do see emission from Crab we can constraint  magnetospheric \Lf\ $\gamma_0$.  Detection  of Crab pulses  at highest frequencies  of $\sim $ 50 GHz imposes  constraints on the properties of particles  accelerating {\it in the \ms}, and the wind properties.

Observations of Crab Nebula require $\kappa \geq 10^5$ for high energy emission \citep{kennel_84,Hibschman,2020ApJ...896..147L}, and even more for radio 
\citep{1977SvA....21..371S,1999A&A...346L..49A}. Models of pulsar high energy emission arising due to Inverse Compton scattering 
\citep{2012ApJ...754...33L,2013MNRAS.431.2580L} also require high multiplicities $\kappa \geq 10^6$, though in that case it's the local/instantaneous   multiplicities, while the estimates for the  Crab Nebula infer average multiplicity.

Single pulses from Crab have been seen at $\sim 50$ GHz \citep{2016ApJ...833...47H}.  Transparency at these frequencies  impose the toughest constraints, as we discuss next.

If expressed in terms of the  surface fields and the spin (known for Crab)  the cyclotron  resonance  occurs at
\be
\frac{ r_{res}}{R_{LC}} =  2 \frac{ B_{NS} R_{NS}^3 \Omega^3}{ m_e c^4 \omega}=
5 
\times 10^3 \nu_9^{-1} 
\label{rres} 
\ee
So, $100$ MHz  should be absorbed at $ 5 
\times 10^4$ \LC\ radii, while $50 GHz$ at  $\sim 100$. 
 \citep[Condition ${\om'} \leq {\om_B'}$ also implies that induced scattering  in the wind ][is suppressed at $r \leq  r_{res} $]{1978MNRAS.185..297W,1992ApJ...395..553S}

Using (\ref{kappa}) and (\ref{15}) the resonant  optical depth is 
\ba &&
\tau_{res} = \frac{\pi}{8}  \frac{ \kappa}{\Gamma} 
\nn && 
\Gamma= {\rm max}[\gamma_0, r_B/R_{LC}]
\ea
Thus,  to see pulses at  50 GHz, where $\gamma_0$ is likely $\geq r_B/R_{LC} \sim 100 $,  it is required that $\gamma_0 \geq  \kappa$: parallel \Lf\ within the \ms\ should be larger than the multiplicity parameter of the \cite{GJ} density. It's a hard one to satisfy - most likely non-stationarity  of the flow plays a role.

\subsection{Rotational phase evolution of Stokes parameters}
Above we  assumes a given direction of linear polarization produced in the \ms, and calculates evolution of the  polarization vector on the Poincare sphere. The tracks on the Poincare sphere we compute are for a single given EM signal with given polarization, as function of the retardance - local plasma parameters times the propagation distance. The observed temporal  evolution of the PA is then a convolution of a possibly phase-dependent  emitted  PA (\eg, given by the Rotating Vector Model), and the propagation effects.

\begin{figure}[h!]
\includegraphics[width=0.9\linewidth]{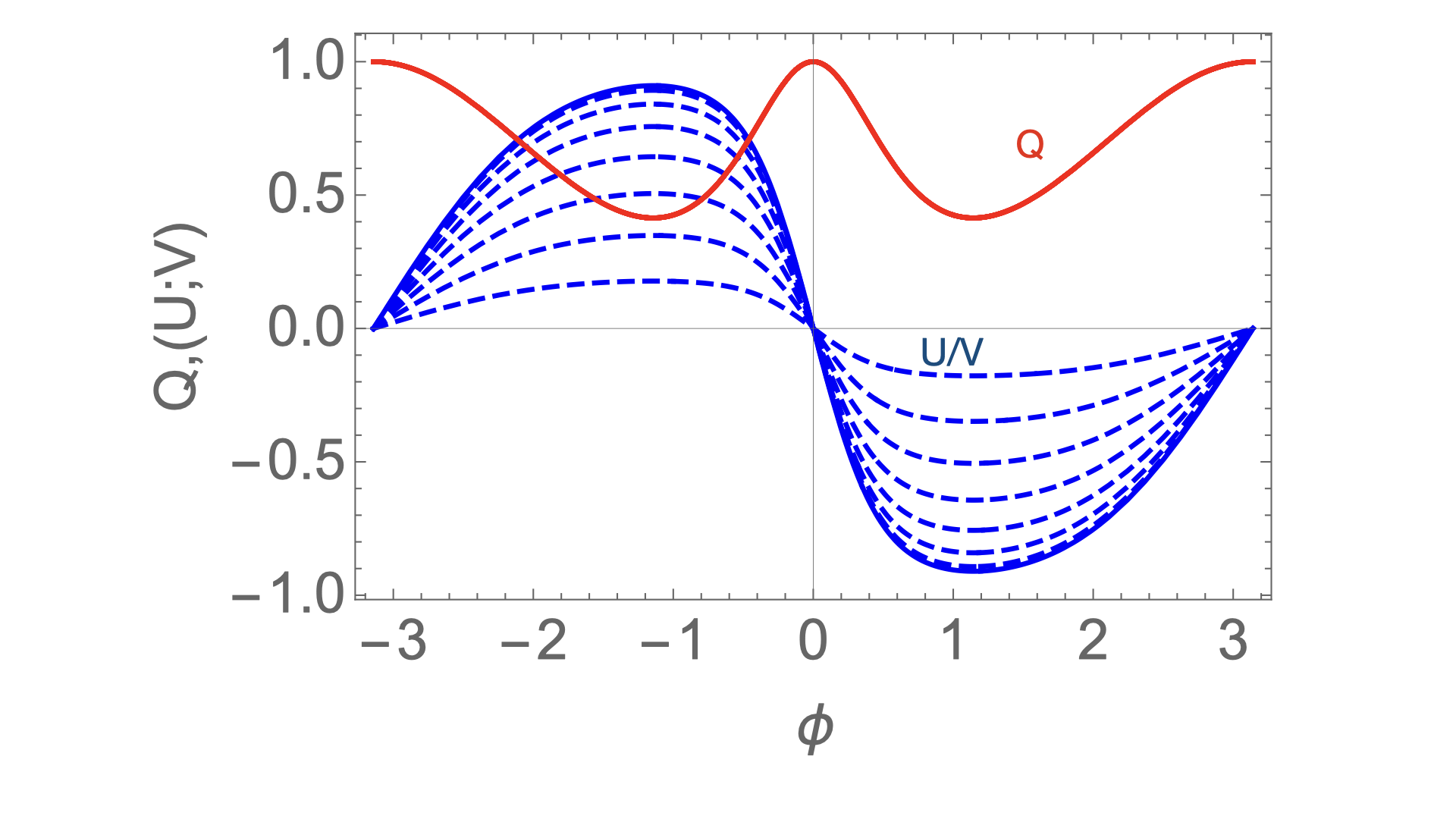}
\caption{Evolution of Stokes parameters as function of rotational phase $\phi$  and different retardance for $\alpha = \pi/8$,  $\theta_{ob} = \pi/4$ assuming that initial phase is given by the rotating vector model. Solid lines correspond to zero retardance (for V=0). Dashed lines are for U-V tracks in  retardance steps of  $\pi/ 16$. U and V curves match each other with a shift of retardance of $\pi/2$ (modulo the sign).}
\label{QUV}
\end {figure}

\begin{figure}[h!]
\includegraphics[width=0.3\linewidth]{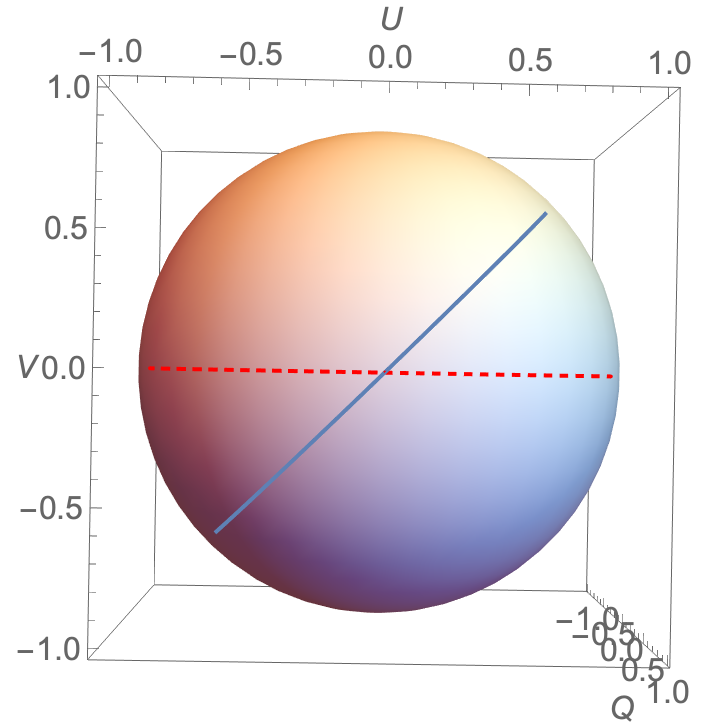}
\includegraphics[width=0.3\linewidth]{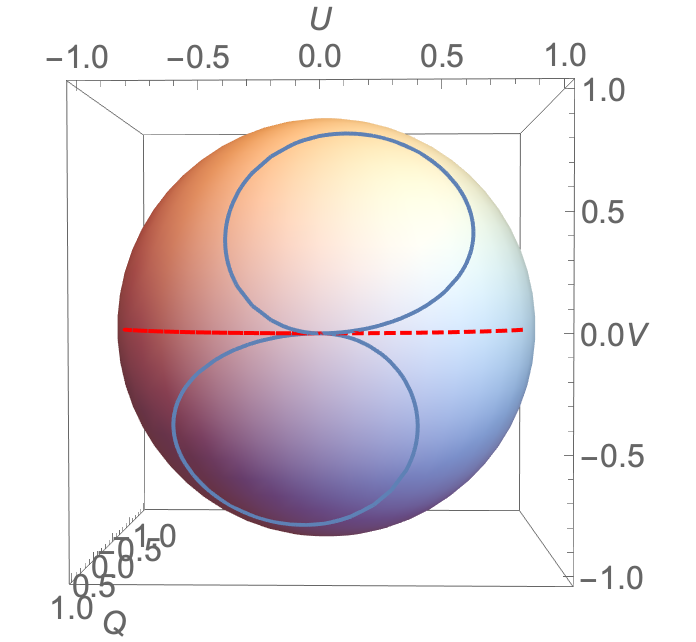}
\includegraphics[width=0.3\linewidth]{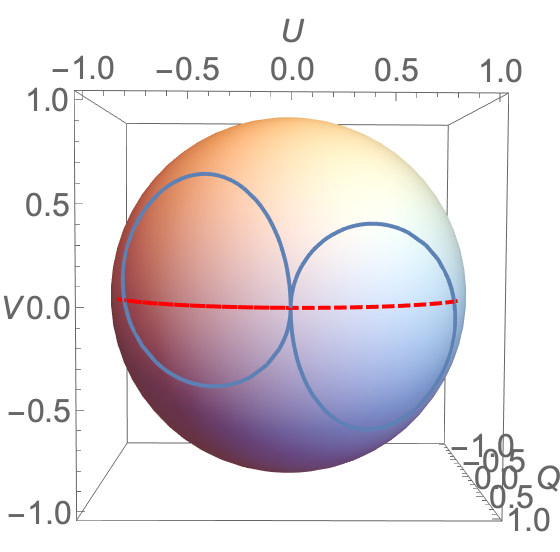}
\caption{Combined effects of RVM and propagation. Pictured are examples of trajectory on the Poincare sphere  for  different  scaling of  the retardace with the phase (\eg due to density variations along the line of sight). If case of no propagation effects the  RVM predicts that the trajectory is a big circle V=0 (dashed red). Left panel: fixed retardace $\delta=\pi/4$ (tilted big circle); Center panel: $\delta=|\phi|$, Right panel: $\delta= \pi/2+ |\phi|$. More complicated dependance of the retardance on the phase would result in correspondingly more complicated trajectories on the  Poincare sphere.}
\label{QUV1}
\end {figure}

For example, using rotating vector model  (RVM) \citep{1969ApL.....3..225R} to produce  $\chi_0$ in Eq. (\ref{chi}), 
\be
\tan \chi _0=\frac{\sin  \alpha \sin  \phi }{\sin  \alpha \cos  \phi 
   \cos \theta _{ob}-\cos  \alpha \sin \theta _{ob}}
   \ee
We can plot evolution of  \{Q,U,V\} as functions of the retardance and rotational phase $\phi$, Fig. \ref{QUV}-\ref{QUV1}. For fixed retardance (left panel in Fig. \ref{QUV1}) tracks on the Poincare sphere may be used to determine absolute position of the projection of the axis of  rotation on the plane of the sky - Q should remain fixed.

We stress that  RVM may not be applicable to magnetar/FRB radio emission (and to Crab pulsar). RVM assumes that global dipole structure of field lines determines local polarization properties of the emitted radiation. In magnetars and FRBs the structure is expected to be highly-non-dipolar, Solar-like \citep{2002ApJ...580L..65L,2015MNRAS.447.1407L,2021ApJ...922..166L}.
Another complication may come from the fact that for oblique rotators the plasma density along the line of sight (and hence the retardance at each moment) may be phase-dependent.

\section{Discussion}

We discuss the properties of polarization transfer in the near wind regions of magnetars, presumed {\it loci} of FRBs; magnetospheric model of radio emission from magnetars and FRBs \citep{2002ApJ...580L..65L,2013arXiv1307.4924P,2020arXiv200505093L,2021ApJ...922..166L} is assumed.
We point out the importance of wave propagation in the inner parts of magnetar's winds. 
Qualitatively, a phase shift between the X and O components of the order of a wavelength, $\sim $ centimeters,   leads to large changes of the polarization properties.

The  birefringent  pair-symmetric plasma   works a wave retarder, periodically converting Stokes $U$ into Stokes $V$ (in a properly defined frame, where one of the axis is aligned  with \Bf).  Unlike the case of  conventional Faraday conversion, here Q-U oscillate in phase (since the Q-U separation depends on the coordinate frame chosen).

The model  offers explanations for  (i)  large circular  polarization component observed in FRBs,  with   right-left switching; (ii) large RM, with possible sign changes (if the observed PA change against frequency is less than 1 rad in the observed frequency range); (iii) time-depend variable polarization. Relatively  dense and  slow wind is needed - the corresponding effect in  regular  pulsars is small. 

The present model  offers a way to produce  large circular polarization and large RM, {\it both  with changing sign} in FRB 20190520B \citep{2022arXiv220308151D}.
 Relatively dense plasma is required -  in regular pulsars this effect is less important that in magnetars (which are expected to produce denser winds).
Pulsars clearly do not show such wild polarization behavior.  The rotating vector model \citep{1969ApL.....3..225R}, that neglects all the propagation effects, does account for many pulsar PA profiles (though there are many exceptions when it's not: e.g. in Crab).  Relation (\ref{RM1}) gives the simplest estimate of the PA effect in the wind. The propagation effects are not important in regular pulsars, with   the low plasma density  and high $\gamma_0$  (the \cite{GJ} density   is much smaller that what is expected in magnetars \citep{tlk,2007ApJ...657..967B}).

The main prediction of the model is that scaling of the PA with frequency {\it  may}  deviate from the conventional  RM with $ \chi \propto \lambda^2$. The reverse is not true: the model  {\it does} allow for $\chi \propto \lambda^2$ scaling, especially at small retardance $\delta \ll 1$.  
There are observational hints, the most interesting analysis is by \cite{2019MNRAS.486.3636P}, their Fig. 8. They demonstrate that {\it linear} relation $\chi \propto \lambda$ is consistent with data. 

Most observational works typically {\it assumes}  a regular RM  scaling: in fact,  the frequency behavior of  polarized components must be fitted independently \citep[\eg][]{2022arXiv220410816K}.

If the system of coordinates  is aligned  with the projection of the spin axis on the plane of the sky then 
during propagation through symmetric  pair wind the Stokes Q remains constant while U and V experience periodic oscillations. The separation between Q and U depends on the chosen system of coordinates, $\times $ versus $ +$. Since the expected track on the Poincare sphere  involves only U-V oscillations, this could be used to determine (to $90^\circ$ uncertainty)   the projection of the pulsar/magnetar spin on the plane of the sky. For example, constant position on the Poincare sphere with $V=0$ may imply that this is a pure Stokes' Q (alternatively explanation would be that propagation effects are not important while the   emitted PA is constant.)

Finally, magnetars show variations of activity on time scales from days to months \citep{2017ARA&A..55..261K}. It is expected that winds are similarly variable. 
 Variations of the density of the wind will lead to medium-to-long  time scale variations of the polarization properties.

\section{ACKNOWLEDGEMENTS}

This work had been supported by NASA grants 80NSSC17K0757 and 80NSSC20K0910, NSF grants 1903332 and 1908590.
I would like to thank Vasily Beskin, Andrei Gruzinov, Marcus Lower, Yuri Levin, Yuri Lyubarski,  Kiyoshi Masui, Donald Melrose, Alexandre Philippov, Eric Poisson  and Louise  Willingale for discussions.

\section{DATA AVAILABILITY}
The data underlying this article will be shared on reasonable request to the corresponding author.

 \bibliographystyle{apj} 
 \bibliography{/Users/maxim/Home/Research/BibTex}

\begin{thebibliography}{59}
\expandafter\ifx\csname natexlab\endcsname\relax\def\natexlab#1{#1}\fi

\bibitem[{{Arons} \& {Barnard}(1986)}]{AronsBarnard86}
{Arons}, J., \& {Barnard}, J.~J. 1986, \apj, 302, 120

\bibitem[{{Arons} \& {Scharlemann}(1979)}]{1979ApJ...231..854A}
{Arons}, J., \& {Scharlemann}, E.~T. 1979, \apj, 231, 854

\bibitem[{{Atoyan}(1999)}]{1999A&A...346L..49A}
{Atoyan}, A.~M. 1999, \aap, 346, L49

\bibitem[{{Barnard}(1986)}]{1986ApJ...303..280B}
{Barnard}, J.~J. 1986, \apj, 303, 280

\bibitem[{{Beloborodov}(2013)}]{2013ApJ...777..114B}
{Beloborodov}, A.~M. 2013, \apj, 777, 114

\bibitem[{{Beloborodov} \& {Thompson}(2007)}]{2007ApJ...657..967B}
{Beloborodov}, A.~M., \& {Thompson}, C. 2007, \apj, 657, 967

\bibitem[{{Beskin} \& {Philippov}(2012)}]{2012MNRAS.425..814B}
{Beskin}, V.~S., \& {Philippov}, A.~A. 2012, \mnras, 425, 814

\bibitem[{{Born} \& {Wolf}(1980)}]{1980poet.book.....B}
{Born}, M., \& {Wolf}, E. 1980, {Principles of Optics Electromagnetic Theory of
  Propagation, Interference and Diffraction of Light}

\bibitem[{{Budden}(1952)}]{1952RSPSA.215..215B}
{Budden}, K.~G. 1952, Proceedings of the Royal Society of London Series A, 215,
  215

\bibitem[{{Caleb} {et~al.}(2019){Caleb}, {van Straten}, {Keane}, {Jameson},
  {Bailes}, {Barr}, {Flynn}, {Ilie}, {Petroff}, {Rogers}, {Stappers},
  {Venkatraman Krishnan}, \& {Weltevrede}}]{2019MNRAS.487.1191C}
{Caleb}, M., {van Straten}, W., {Keane}, E.~F., {Jameson}, A., {Bailes}, M.,
  {Barr}, E.~D., {Flynn}, C., {Ilie}, C.~D., {Petroff}, E., {Rogers}, A.,
  {Stappers}, B.~W., {Venkatraman Krishnan}, V., \& {Weltevrede}, P. 2019,
  \mnras, 487, 1191

\bibitem[{{Cheng} \& {Ruderman}(1979)}]{1979ApJ...229..348C}
{Cheng}, A.~F., \& {Ruderman}, M.~A. 1979, \apj, 229, 348

\bibitem[{{Dai} {et~al.}(2022){Dai}, {Feng}, {Yang}, {Zhang}, {Li}, {Niu},
  {Wang}, {Xue}, {Zhang}, {Burke-Spolaor}, {Law}, {Lynch}, {Connor},
  {Anna-Thomas}, {Zhang}, {Duan}, {Yao}, {Tsai}, {Zhu}, {Cruces}, {Hobbs},
  {Miao}, {Niu}, {Filipovic}, \& {Zhu}}]{2022arXiv220308151D}
{Dai}, S., {Feng}, Y., {Yang}, Y.~P., {Zhang}, Y.~K., {Li}, D., {Niu}, C.~H.,
  {Wang}, P., {Xue}, M.~Y., {Zhang}, B., {Burke-Spolaor}, S., {Law}, C.~J.,
  {Lynch}, R.~S., {Connor}, L., {Anna-Thomas}, R., {Zhang}, L., {Duan}, R.,
  {Yao}, J.~M., {Tsai}, C.~W., {Zhu}, W.~W., {Cruces}, M., {Hobbs}, G., {Miao},
  C.~C., {Niu}, J.~R., {Filipovic}, M.~D., \& {Zhu}, S.~Q. 2022, arXiv
  e-prints, arXiv:2203.08151

\bibitem[{{Ginzburg} \& {Syrovatskii}(1965)}]{1965ARA&A...3..297G}
{Ginzburg}, V.~L., \& {Syrovatskii}, S.~I. 1965, \araa, 3, 297

\bibitem[{{Goldreich} \& {Julian}(1969)}]{GJ}
{Goldreich}, P., \& {Julian}, W.~H. 1969, \apj, 157, 869

\bibitem[{{Goldreich} \& {Julian}(1970)}]{1970ApJ...160..971G}
---. 1970, \apj, 160, 971

\bibitem[{{Gruzinov} \& {Levin}(2019)}]{2019ApJ...876...74G}
{Gruzinov}, A., \& {Levin}, Y. 2019, \apj, 876, 74

\bibitem[{{Hankins} {et~al.}(2016){Hankins}, {Eilek}, \&
  {Jones}}]{2016ApJ...833...47H}
{Hankins}, T.~H., {Eilek}, J.~A., \& {Jones}, G. 2016, \apj, 833, 47

\bibitem[{{Hibschman} \& {Arons}(2001)}]{Hibschman}
{Hibschman}, J.~A., \& {Arons}, J. 2001, \apj, 560, 871

\bibitem[{{Kaspi} \& {Beloborodov}(2017)}]{2017ARA&A..55..261K}
{Kaspi}, V.~M., \& {Beloborodov}, A.~M. 2017, \araa, 55, 261

\bibitem[{{Kazbegi} {et~al.}(1991{\natexlab{a}}){Kazbegi}, {Machabeli}, \&
  {Melikidze}}]{1991MNRAS.253..377K}
{Kazbegi}, A.~Z., {Machabeli}, G.~Z., \& {Melikidze}, G.~I. 1991{\natexlab{a}},
  \mnras, 253, 377

\bibitem[{{Kazbegi} {et~al.}(1991{\natexlab{b}}){Kazbegi}, {Machabeli},
  {Melikidze}, \& {Smirnova}}]{Kaz91}
{Kazbegi}, A.~Z., {Machabeli}, G.~Z., {Melikidze}, G.~I., \& {Smirnova}, T.~V.
  1991{\natexlab{b}}, Astrophysics, 34, 234

\bibitem[{{Kennel} \& {Coroniti}(1984)}]{kennel_84}
{Kennel}, C.~F., \& {Coroniti}, F.~V. 1984, \apj, 283, 710

\bibitem[{{Kennett} \& {Melrose}(1998)}]{1998PASA...15..211K}
{Kennett}, M., \& {Melrose}, D. 1998, \pasa, 15, 211

\bibitem[{{Kumar} {et~al.}(2022){Kumar}, {Shannon}, {Lower}, {Deller}, \&
  {Prochaska}}]{2022arXiv220410816K}
{Kumar}, P., {Shannon}, R.~M., {Lower}, M.~E., {Deller}, A.~T., \& {Prochaska},
  J.~X. 2022, arXiv e-prints, arXiv:2204.10816

\bibitem[{{Landau} \& {Lifshitz}(1960)}]{LLVIII}
{Landau}, L.~D., \& {Lifshitz}, E.~M. 1960, {Electrodynamics of continuous
  media} (Energy Conversion Management)

\bibitem[{{Landau} \& {Lifshitz}(1975)}]{LLII}
---. 1975, {The classical theory of fields}

\bibitem[{{Luo} {et~al.}(2020){Luo}, {Lyutikov}, {Temim}, \&
  {Comisso}}]{2020ApJ...896..147L}
{Luo}, Y., {Lyutikov}, M., {Temim}, T., \& {Comisso}, L. 2020, \apj, 896, 147

\bibitem[{{Lyutikov}(1999)}]{1999JPlPh..62...65L}
{Lyutikov}, M. 1999, Journal of Plasma Physics, 62, 65

\bibitem[{{Lyutikov}(2002)}]{2002ApJ...580L..65L}
---. 2002, \apjl, 580, L65

\bibitem[{{Lyutikov}(2007)}]{2007MNRAS.381.1190L}
---. 2007, \mnras, 381, 1190

\bibitem[{{Lyutikov}(2013)}]{2013MNRAS.431.2580L}
---. 2013, \mnras, 431, 2580

\bibitem[{{Lyutikov}(2015)}]{2015MNRAS.447.1407L}
---. 2015, \mnras, 447, 1407

\bibitem[{{Lyutikov}(2021)}]{2021ApJ...922..166L}
---. 2021, \apj, 922, 166

\bibitem[{{Lyutikov} \& {Gavriil}(2006)}]{lg06}
{Lyutikov}, M., \& {Gavriil}, F.~P. 2006, MNRAS, 368, 690

\bibitem[{{Lyutikov} {et~al.}(2012){Lyutikov}, {Otte}, \&
  {McCann}}]{2012ApJ...754...33L}
{Lyutikov}, M., {Otte}, N., \& {McCann}, A. 2012, \apj, 754, 33

\bibitem[{{Lyutikov} \& {Popov}(2020)}]{2020arXiv200505093L}
{Lyutikov}, M., \& {Popov}, S. 2020, arXiv e-prints, arXiv:2005.05093

\bibitem[{{Masui} {et~al.}(2015){Masui}, {Lin}, {Sievers}, {Anderson}, {Chang},
  {Chen}, {Ganguly}, {Jarvis}, {Kuo}, {Li}, {Liao}, {McLaughlin}, {Pen},
  {Peterson}, {Roman}, {Timbie}, {Voytek}, \& {Yadav}}]{2015Natur.528..523M}
{Masui}, K., {Lin}, H.-H., {Sievers}, J., {Anderson}, C.~J., {Chang}, T.-C.,
  {Chen}, X., {Ganguly}, A., {Jarvis}, M., {Kuo}, C.-Y., {Li}, Y.-C., {Liao},
  Y.-W., {McLaughlin}, M., {Pen}, U.-L., {Peterson}, J.~B., {Roman}, A.,
  {Timbie}, P.~T., {Voytek}, T., \& {Yadav}, J.~K. 2015, \nat, 528, 523

\bibitem[{{Melrose}(1997)}]{1997PhRvE..56.3527M}
{Melrose}, D.~B. 1997, \pre, 56, 3527

\bibitem[{{Michel}(1969)}]{1969ApJ...158..727M}
{Michel}, F.~C. 1969, \apj, 158, 727

\bibitem[{{Michel}(1973{\natexlab{a}})}]{Michel73}
---. 1973{\natexlab{a}}, \apj, 180, 207

\bibitem[{{Michel}(1973{\natexlab{b}})}]{1973ApJ...180L.133M}
---. 1973{\natexlab{b}}, \apjl, 180, L133

\bibitem[{{Michilli} {et~al.}(2018){Michilli}, {Seymour}, {Hessels}, {Spitler},
  {Gajjar}, {Archibald}, {Bower}, {Chatterjee}, {Cordes}, {Gourdji}, {Heald},
  {Kaspi}, {Law}, {Sobey}, {Adams}, {Bassa}, {Bogdanov}, {Brinkman},
  {Demorest}, {Fernandez}, {Hellbourg}, {Lazio}, {Lynch}, {Maddox}, {Marcote},
  {McLaughlin}, {Paragi}, {Ransom}, {Scholz}, {Siemion}, {Tendulkar}, {van
  Rooy}, {Wharton}, \& {Whitlow}}]{2018Natur.553..182M}
{Michilli}, D., {Seymour}, A., {Hessels}, J.~W.~T., {Spitler}, L.~G., {Gajjar},
  V., {Archibald}, A.~M., {Bower}, G.~C., {Chatterjee}, S., {Cordes}, J.~M.,
  {Gourdji}, K., {Heald}, G.~H., {Kaspi}, V.~M., {Law}, C.~J., {Sobey}, C.,
  {Adams}, E.~A.~K., {Bassa}, C.~G., {Bogdanov}, S., {Brinkman}, C.,
  {Demorest}, P., {Fernandez}, F., {Hellbourg}, G., {Lazio}, T.~J.~W., {Lynch},
  R.~S., {Maddox}, N., {Marcote}, B., {McLaughlin}, M.~A., {Paragi}, Z.,
  {Ransom}, S.~M., {Scholz}, P., {Siemion}, A.~P.~V., {Tendulkar}, S.~P., {van
  Rooy}, P., {Wharton}, R.~S., \& {Whitlow}, D. 2018, \nat, 553, 182

\bibitem[{{Petroff} {et~al.}(2019){Petroff}, {Hessels}, \&
  {Lorimer}}]{2019A&ARv..27....4P}
{Petroff}, E., {Hessels}, J.~W.~T., \& {Lorimer}, D.~R. 2019, \aapr, 27, 4

\bibitem[{{Petrova} \& {Lyubarskii}(2000)}]{2000A&A...355.1168P}
{Petrova}, S.~A., \& {Lyubarskii}, Y.~E. 2000, \aap, 355, 1168

\bibitem[{{Popov} \& {Postnov}(2013)}]{2013arXiv1307.4924P}
{Popov}, S.~B., \& {Postnov}, K.~A. 2013, arXiv e-prints, arXiv:1307.4924

\bibitem[{{Price} {et~al.}(2019){Price}, {Foster}, {Geyer}, {van Straten},
  {Gajjar}, {Hellbourg}, {Karastergiou}, {Keane}, {Siemion}, {Arcavi}, {Bhat},
  {Caleb}, {Chang}, {Croft}, {DeBoer}, {de Pater}, {Drew}, {Enriquez}, {Farah},
  {Gizani}, {Green}, {Isaacson}, {Hickish}, {Jameson}, {Lebofsky}, {MacMahon},
  {M{\"o}ller}, {Onken}, {Petroff}, {Werthimer}, {Wolf}, {Worden}, \&
  {Zhang}}]{2019MNRAS.486.3636P}
{Price}, D.~C., {Foster}, G., {Geyer}, M., {van Straten}, W., {Gajjar}, V.,
  {Hellbourg}, G., {Karastergiou}, A., {Keane}, E.~F., {Siemion}, A.~P.~V.,
  {Arcavi}, I., {Bhat}, R., {Caleb}, M., {Chang}, S.~W., {Croft}, S., {DeBoer},
  D., {de Pater}, I., {Drew}, J., {Enriquez}, J.~E., {Farah}, W., {Gizani}, N.,
  {Green}, J.~A., {Isaacson}, H., {Hickish}, J., {Jameson}, A., {Lebofsky}, M.,
  {MacMahon}, D.~H.~E., {M{\"o}ller}, A., {Onken}, C.~A., {Petroff}, E.,
  {Werthimer}, D., {Wolf}, C., {Worden}, S.~P., \& {Zhang}, Y.~G. 2019, \mnras,
  486, 3636

\bibitem[{{Radhakrishnan} \& {Cooke}(1969)}]{1969ApL.....3..225R}
{Radhakrishnan}, V., \& {Cooke}, D.~J. 1969, \aplett, 3, 225

\bibitem[{{Ravi} {et~al.}(2016){Ravi}, {Shannon}, {Bailes}, {Bannister},
  {Bhandari}, {Bhat}, {Burke-Spolaor}, {Caleb}, {Flynn}, {Jameson}, {Johnston},
  {Keane}, {Kerr}, {Tiburzi}, {Tuntsov}, \& {Vedantham}}]{2016Sci...354.1249R}
{Ravi}, V., {Shannon}, R.~M., {Bailes}, M., {Bannister}, K., {Bhandari}, S.,
  {Bhat}, N.~D.~R., {Burke-Spolaor}, S., {Caleb}, M., {Flynn}, C., {Jameson},
  A., {Johnston}, S., {Keane}, E.~F., {Kerr}, M., {Tiburzi}, C., {Tuntsov},
  A.~V., \& {Vedantham}, H.~K. 2016, Science, 354, 1249

\bibitem[{{Ruderman} \& {Sutherland}(1975)}]{1975ApJ...196...51R}
{Ruderman}, M.~A., \& {Sutherland}, P.~G. 1975, \apj, 196, 51

\bibitem[{{Sazonov}(1969)}]{1969SvA....13..396S}
{Sazonov}, V.~N. 1969, \sovast, 13, 396

\bibitem[{{Shklovskii}(1977)}]{1977SvA....21..371S}
{Shklovskii}, I.~S. 1977, \sovast, 21, 371

\bibitem[{{Sincell} \& {Krolik}(1992)}]{1992ApJ...395..553S}
{Sincell}, M.~W., \& {Krolik}, J.~H. 1992, \apj, 395, 553

\bibitem[{{Thompson} {et~al.}(1994){Thompson}, {Blandford}, {Evans}, \&
  {Phinney}}]{1994ApJ...422..304T}
{Thompson}, C., {Blandford}, R.~D., {Evans}, C.~R., \& {Phinney}, E.~S. 1994,
  \apj, 422, 304

\bibitem[{{Thompson} {et~al.}(2002){Thompson}, {Lyutikov}, \& {Kulkarni}}]{tlk}
{Thompson}, C., {Lyutikov}, M., \& {Kulkarni}, S.~R. 2002, \apj, 574, 332

\bibitem[{{Timokhin}(2010)}]{2010MNRAS.408.2092T}
{Timokhin}, A.~N. 2010, \mnras, 408, 2092

\bibitem[{{Vedantham} \& {Ravi}(2019)}]{2019MNRAS.485L..78V}
{Vedantham}, H.~K., \& {Ravi}, V. 2019, \mnras, 485, L78

\bibitem[{{Wang} {et~al.}(2010){Wang}, {Lai}, \& {Han}}]{2010MNRAS.403..569W}
{Wang}, C., {Lai}, D., \& {Han}, J. 2010, \mnras, 403, 569

\bibitem[{{Wilson} \& {Rees}(1978)}]{1978MNRAS.185..297W}
{Wilson}, D.~B., \& {Rees}, M.~J. 1978, \mnras, 185, 297

\bibitem[{{Zheleznyakov}(1996)}]{1996ASSL..204.....Z}
{Zheleznyakov}, V.~V. 1996, {Radiation in Astrophysical Plasmas}, Vol. 204

\end{thebibliography}

 \end{document}